\documentclass[12pt]{iopart}

\usepackage{verbatim} 
\usepackage{multirow}

\usepackage{graphicx}
\usepackage{graphics}
\usepackage{subfigure}
\usepackage{array}

\usepackage{ulem}

\usepackage{epsfig}

\begin{document}

\title{Weighing of trapped ion crystals and its applications}

\author{Kevin Sheridan and Matthias Keller}

\address{Department of Physics and Astronomy, University of Sussex, Falmer, BN1 9QH, UK}
\ead{kts20@sussex.ac.uk}
\begin{abstract}
We have developed a novel scheme to measure the secular motion of trapped ions. Employing pulsed excitation and analysis of the fluorescence of laser cooled ions, we have measured the centre-of-mass mode frequency of single as well as entire ion crystals with a frequency precision better than $5\cdot 10^{-4}$ within an interrogation time on the order of seconds, limited only by the fluorescence collection efficiency and the background noise. We have used this method to measure the mass of ions and observed charge exchange collisions between trapped calcium isotopes.

\end{abstract}

\submitto{\NJP}
\maketitle

\section{Introduction}
\label{sec:intro}

The precise measurement of the secular motion of ions in a Paul trap is an important tool in modern atomic physics. In many applications, most notably trapped ion quantum information processing (QIP) \cite{Ekert,Haffner,Leibfried2} and metrology \cite{Roos,Rosenband,Margolis,Stenger} the motion of the ions must be cooled close to the motional ground state, which requires the precise knowledge of the motional spectrum of the trapped ions. 
Furthermore, it is crucial for ion-trap quantum information experiments to enable high fidelity quantum gates \cite{Leibfried3,Cirac}.
Recently, the measurement of the individual secular frequencies of ions in large multi-component ion crystals has been exploited to investigate the photo-dissociation of large molecules \cite{Offenberg} and to perform high resolution spectroscopy of molecular ions \cite{Roth}. These novel applications make it especially interesting to investigate innovative ways of performing secular frequency measurements quickly and with high precision.

Often the secular frequency of ions is measured by continuous excitation of the ions' motion and the simultaneous observation of their fluorescence level \cite{Welling} or the phase between the ions' oscillation and the excitation \cite{Drewsen}.
In multi-component ion crystals, the excitation of the secular motion of one ion species may heat the entire crystal which results in a decrease in fluorescence \cite{Baba}. Side-band spectroscopy \cite{Raab, Monroe}, which is used routinely in QIP and high resolution spectroscopy with trapped ions, can also be employed to measure the secular frequency of ions.

Instead of utilising the change in the fluorescence level, the induced modulation of the fluorescence can be used to extract a spectroscopic signal. The ion's oscillation leads to a modulation of its fluorescence which can be measured by correlating the fluorescence with the external modulation \cite{Blumel,Dholakia,Berkeland,Bollinger} or by direct Fourier transformation of its fluorescence \cite{Gerlich}. Alternatively, the second-order correlation function (g$^{(2)}$-function) of the fluorescence can be measured \cite{Diedrich}. D Rotter et al.\ \cite{Rotter} have recently measured the motional spectrum of ions through the g$^{(2)}$-function by measuring the ions' fluorescence which is partly retro-reflected by an additional mirror.

In this paper we present a novel method to measure the centre-of-mass (COM) mode frequency which is suitable for a wide range of ion crystals reaching from single ions to several hundred ions. We employ the Fourier transform of the auto-correlation of the ions' fluorescence intensity while exciting their motion by a pulsed electric field. The resolution of this method is governed by the motional damping due to laser cooling and the signal-to-noise ratio (SNR) of the fluorescence. We demonstrate a resolution of better than 100~Hz with an interrogation time in the order of seconds. The increase in the ion's average temperature is less than 3~mK and the method is shown to work well even with large ion crystals. The technique we present is unique in the fact that it provides an accurate and fast real-time secular frequency measurement without the need for modifications to the standard laser cooling scheme, i.e.\ only the fluorescence during Doppler cooling is analysed. Owing to its non-invasive nature, it can be employed in a diverse range of novel applications.

In the first part of this paper we discuss the principles of this method and the experimental set-up. The characterisation is then presented, where the secular frequency of a single trapped ion is measured and the results are compared to a model of the system. In the last section, we present some of the possible applications of this method.  For a short mixed ion string we measure the axial COM-mode frequency of the different ion configurations and extract the mass of a `dark ion' embedded in a short ion string. We demonstrate that this method can be utilised to measure the COM-mode frequency of three-dimensional ion crystals with several hundred ions without deteriorating their structure. In addition, we employ the high resolution of this method to measure the charge exchange between $^{44}$Ca-ions and neutral calcium atoms by observing the change in the COM-mode frequency of a small crystal.

\section{Measurement principle}
\label{sec:measurementPrinciple}
The measurement principle is based on the detection of the Doppler induced modulation of the ion's fluorescence due to its motion. In contrast to the other methods described above, the motion is excited by short voltage pulses and detected through the auto-correlation of the ion's fluorescence. The secular frequency in then determined by Fourier transforming the auto-correlation.

In the weak binding regime, the driven oscillatory motion of an ion is converted into a modulation of its fluorescence due to the Doppler effect. For small velocities $\vec{v}$ the amplitude modulation of the laser induced fluorescence $\Delta \mbox{Fl}$ can be expressed as:
\begin{equation}
\Delta \mbox{Fl} = \Gamma\left.\frac{d\rho_{ee}}{d\delta}\right|_\delta \vec{k}\cdot\vec{v},
\label{equ:fluoresMod1}
\end{equation}
with the detuning between the laser and the atomic resonance $\delta$, the laser wave vector $\vec{k}$, the spontaneous decay rate $\Gamma$ and the excited state population $\rho_{ee}$.
In the simple case of an ion with a closed two-level system this becomes:
\begin{equation}
\Delta \mbox{Fl} = \frac{\Gamma s \delta}{\left[\left(1+s\right)\Gamma^2+\delta^2\right]^2}\, \vec{k}\cdot\vec{v},
\label{equ:fluoresMod2}
\end{equation}
with the transition saturation parameter $s$.
Thus the gradient is only determined by the saturation parameter, the laser detuning and the spontaneous decay rate $\Gamma$.


Instead of using a monochromatic excitation, we employ periodic broadband pulses to access a wide range of secular frequencies simultaneously.
The motion is driven by these short voltage pulses which are applied to one of the trap dc-electrodes. For an excitation with period $T$ and a pulse length $\tau$, the pulse train can be expressed as:
\begin{equation}
U(t) = \frac{U_0\tau}{2T}+\frac{2U_0\tau}{\pi T}\sum^\infty_{n=1}\mbox{sinc}\!\left(\frac{n\pi\tau}{T}\right)\cos\!\left(\frac{2\pi nt}{T}\right).
\label{equ:combspectrum}
\end{equation}
Thus, it consists of a comb-like spectrum with equidistant spacing of 1/$T$, the repetition rate of the excitation, and a sinc shaped envelope which is determined by the pulse width $\tau$. The width of the individual comb peaks can be neglected in most cases.

During each voltage pulse, the trapping potential is instantaneously shifted which leads to the excitation of the COM-mode of the trapped ions. Due to the laser induced damping of the motion, the oscillation amplitude decays and this results in a damped modulation of the ion's fluorescence. 

In order to detect the Doppler induced modulation of the ion's fluorescence, we employ the auto-correlation of the fluorescence measured with a photo-multiplier. For each photon detection event, the detection time is recorded. After the interrogation time, the list of time stamps is then auto-correlated and the spectrum is obtained by its Fourier transform. In contrast to the cross-correlation of the photon detection times with the excitation, the auto-correlation is insensitive to timing jitters of the excitation and, most importantly, it provides a flat background in the spectrum.

\begin{figure}
\begin{center}
\includegraphics[width=0.8\linewidth]{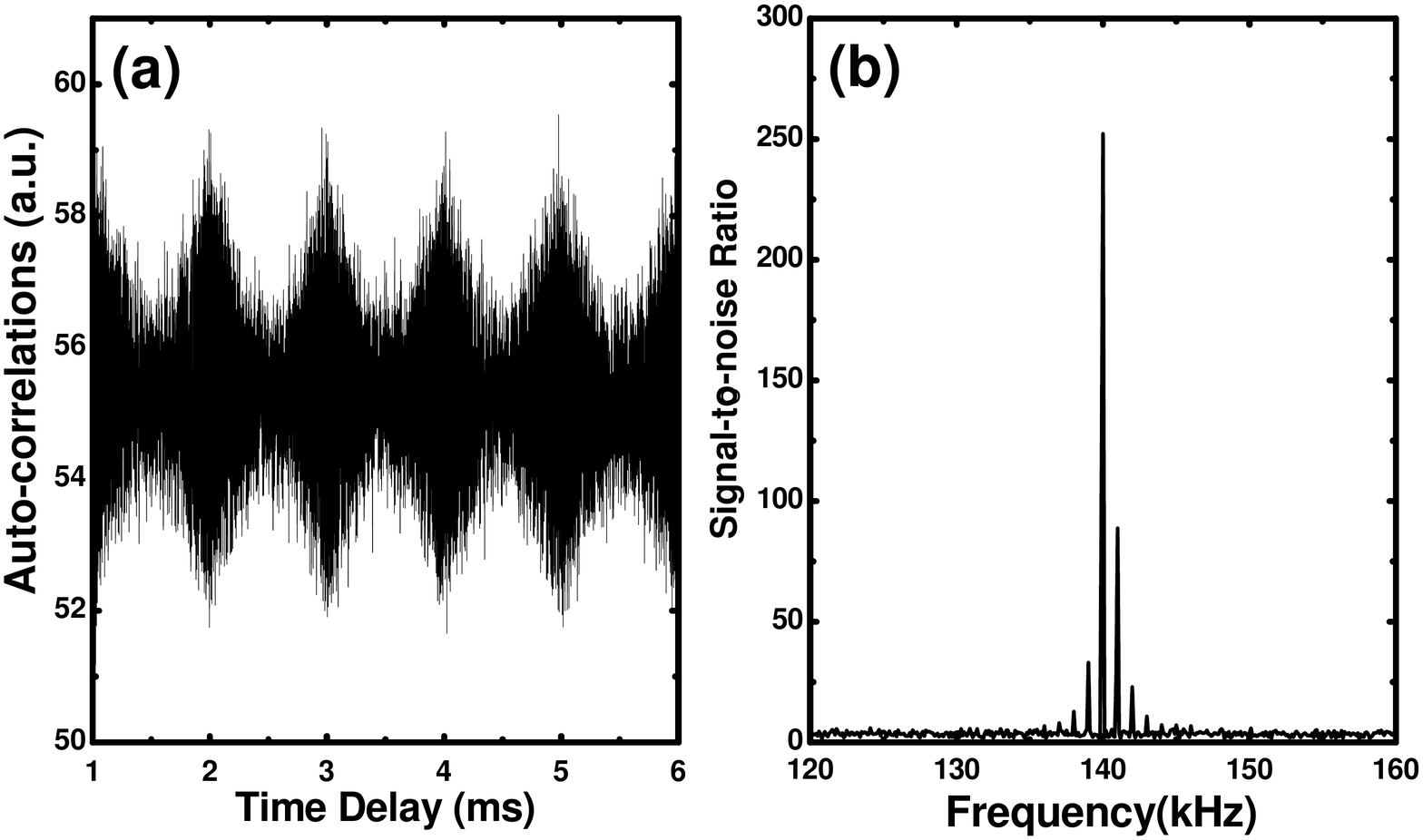}
\caption{Auto-correlation of the arrival times of photons collected from the modulated single ion. The interrogation time is 45 seconds, the driving pulse amplitude is 2.0~V, the repetition rate is 1~kHz and the pulse width is 3~$\mu$s. An axial cooling laser power of 10~$\mu$W is used.}
\label{fig:measurementexample}
\end{center}
\end{figure}

Figure \ref{fig:measurementexample}(a) shows a typical auto-correlation of the ion's fluorescence.
The periodic excitation is clearly visible, with the occurrence of regular maxima in the oscillation of the auto-correlations together with the laser induced damping. Figure \ref{fig:measurementexample}(b) shows the spectrum of the auto-correlation measurement with the high contrast secular resonance. Due to the periodical pulsed excitation described by equation \ref{equ:combspectrum}, the spectrum consist of a comb of lines under a Lorentzian envelope corresponding to the motional resonance. The peak separation reflects the repetition rate of the excitation. By fitting a Lorentzian to the spectrum, the precise secular frequency can be determined.

\section{Experimental set-up}
\label{sec:experiment_setup}

\begin{figure}[h]
\begin{center}
\includegraphics[width=0.8\linewidth]{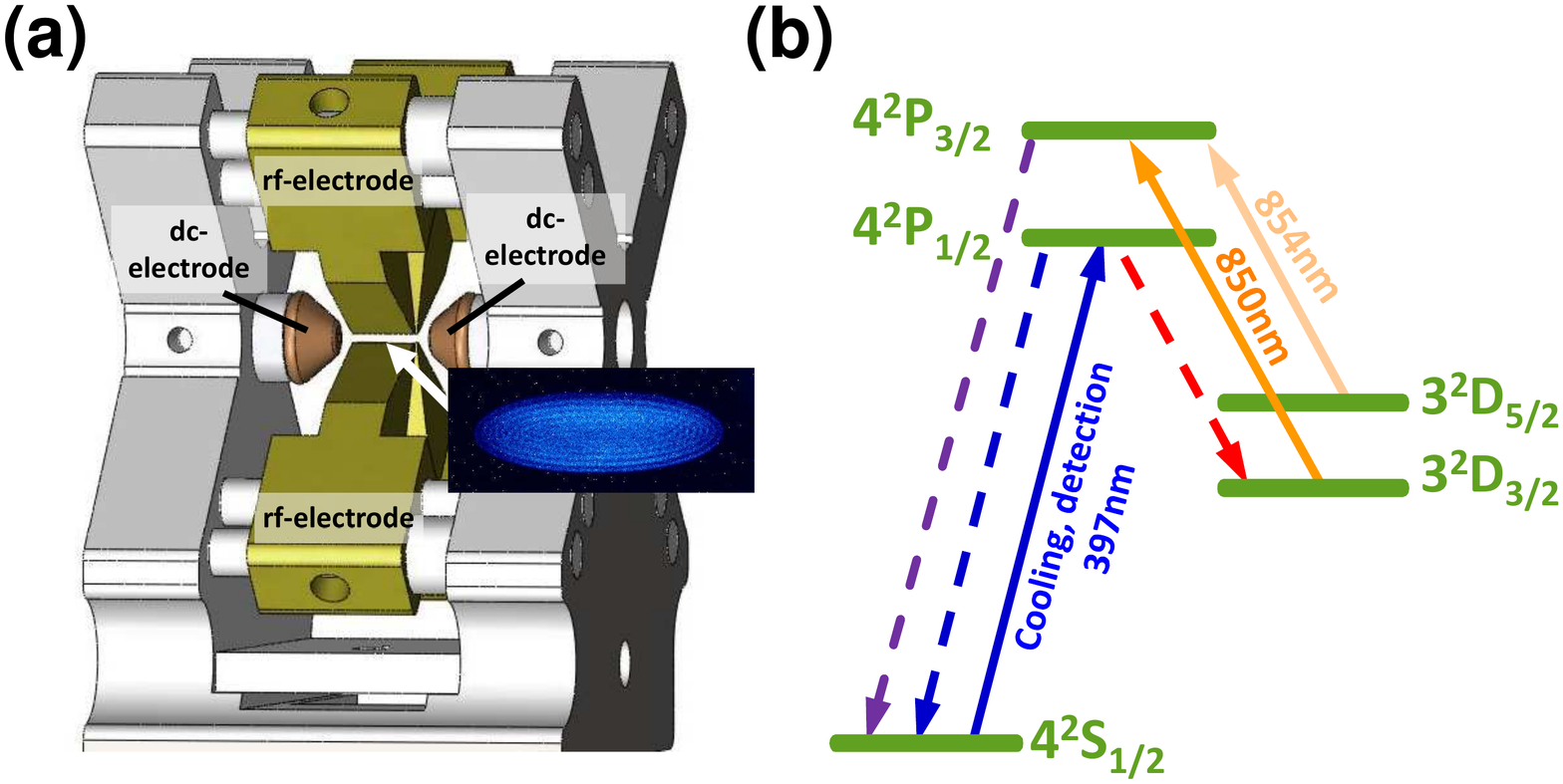}
\caption[]{(a) Schematic of the ion trap. (b) Level scheme of $^{40}$Ca-ion including all lasers required to cool and re-pump the ion (solid arrows). Dashed lines indicate spontaneous emission.}
\label{fig:iontrap}
\end{center}
\end{figure}

In order to investigate this method to detect the COM-mode frequency of ions, $^{40}$Ca-ions in a linear Paul trap are employed. The trap has been designed for the investigation of charge exchange processes and chemical reactions between ions and neutral atoms or molecules and offers good optical access. It consists of four blade shaped rf-electrodes with an ion-electrode distance of 465~$\mu$m and an rf-electrode length of 4~mm which provide the radial confinement of the ions. Positive static potentials applied to two dc-electrodes, separated by 6~mm are used for the axial ion confinement. A schematic of the ion trap is shown in figure \ref{fig:iontrap}(a).  The trap geometry has been designed to be  self-aligning, so that the electrode positions are entirely determined by the machining process. The electrode mount as well as the rf-electrodes are machined out of one block of stainless steel after precision holes have been reamed. After separating the parts and shaping the electrodes, the trap can be easily assembled by inserting precision ceramic dowel pins through the rf-electrodes and into the mount.
The dc-electrodes for the axial confinement have openings in order to provide laser access to the trapped ions exactly along the trap axis.
The rf-electrodes form the capacitor of a LC-circuit with a quality factor of 300 for the trap frequency of 29.579~MHz. This rf-circuit is driven directly by a function generator without additional amplifier. The radial secular frequency of the ions is 1.5~MHz. In the axial direction, a positive potential of up to 500~V on the dc-electrodes leads to an axial secular frequency of 380~kHz.

The $^{40}$Ca-ions are Doppler cooled on the 4S$_{1/2}\!\!\rightarrow $4P$_{1/2}$ transition at 397~nm. In order to avoid populating the meta-stable D$_{3/2}$-state through the decay of the P$_{1/2}$-level, a re-pump laser on the D$_{3/2}\!\!\rightarrow $P$_{3/2}$ transition at 850nm is applied (see figure \ref{fig:iontrap}(b)). Thus, the ion is re-pumped into the ground state through the spontaneous decay of the P$_{3/2}$-state. However, another re-pump laser at 854~nm must be applied in order to avoid optical pumping into the meta-stable D$_{5/2}$-state. This re-pump scheme has the advantage over re-pumping via the P$_{1/2}$-level in that no coherent population trapping with the cooling laser occurs. This creates an effective two-level cooling system. To cool all directions of the ion's motion, one laser beam is aligned to have an angle of 76$^\circ$ with respect to the trap axis. Another cooling laser, together with the re-pumping lasers, is aligned through the holes in the dc-electrodes and is thus well aligned with the trap axis. The powers and frequencies of the re-pump lasers are adjusted to obtain the maximal fluorescence. The ratio of the two cooling lasers is set such that most of the fluorescence is generated by the axial laser but enough radial cooling is provided for good localisation of the ions. This is particularly important for larger ion crystals. In this way the axial secular motion can be efficiently detected as the ions motion and the laser direction are collinear. 

Fluorescence at 397~nm is collected by a microscope lens with x10 magnification and a numerical aperture of 0.22 and divided by a beam splitter so that the CCD and photo-multiplier can be used simultaneously. The detection efficiency of the fluorescence with the photo-multiplier is about 0.1\%.

The trap is loaded through the photoionisation of naturally abundant neutral calcium atoms effusing from a resistively heated oven located below the trap and aligned perpendicular to the trap axis. Photoionisation is a two-photon process with the first photon at 423~nm resonant with the S$_{0}\!\!\rightarrow$P$_{1}$ transition of neutral calcium and the second photon at 375~nm providing the energy necessary to ionise the excited state calcium atoms \cite{Lucas}.
For appropriate trapping parameters single ions, linear ion strings and large three dimensional Coulomb crystals with several hundred ions can be quickly and repeatedly loaded into the trap.

To prepare the secular frequency measurement, the cooling laser is set to the appropriate detuning and the ion's fluorescence is detected with a PMT. At this point the motional excitation of the ion is activated. Positive voltage pulses with a typical duration of several $\mu$s and voltage in the order of a few V are applied to one of the dc-electrodes. The repetition rate of the pulses is typically between 500~Hz and several kHz. A RC circuit is used to superimpose the static dc voltage for the ion's axial confinement with the driving voltage pulses. 

During the measurement the photon detection times are recorded with a time to digital converter (FastComtec 7888.)  After the chosen measurement time, the auto-correlation of the photon detection times is calculated, from which the spectrum is obtained by Fourier transformation. 
The frequency resolution can be improved beyond the repetition rate by fitting a Lorentzian to the auto-correlation spectra. With this, an accuracy of better than 30~Hz has been achieved for a repetition rate of 500~Hz.

We have conducted a detailed numerical analysis which includes the back action of the measurement on the ion's motion. Our model is based on the ion's harmonic motion along the trap axis including the non-linear damping term due to the laser induced radiation pressure of the form $\hbar \vec{k}\Gamma \rho_{ee}\left( \delta = \delta_0+\vec{k}\vec{v}\right)$. The ion's velocity evolution has been obtained from the numerical integration of the ion's equation of motion during the excitation pulse as well as the laser damped oscillation. This in turn is converted into fluorescence through $\mbox{Fl}=\Gamma \rho_{ee}\left( \delta = \delta_0+\vec{k}\vec{v}\right)$. In order to obtain an accurate representation of the expected signal, the ion's motion and its fluorescence has been simulated for a series of several hundred excitation pulses. The numerical fluorescence data has been processed in the same way as the measured data from the experiment. Firstly, the auto-correlation of the fluorescences has been performed followed by the fast Fourier transform of the auto-correlation function. 


\section{Measurement characterisation}
\label{sec:measurement_characterisation}
By thoroughly characterising our measurement scheme the optimal settings of the experimental parameters can be found to achieve the maximal frequency resolution. These parameters can be split into two categories: those associated with the interaction of the radiation with the ion and those related to the excitation of the ion's motion. The parameters of the periodic voltage pulses applied to the dc-electrode strongly impact the ion's motion and thus the amplitude of the auto-correlation spectrum. However, these parameters will not affect the conversion of the ion's oscillation into fluorescence modulation. In contrast, the laser parameters impact this conversion by changing the ion's spectral line shape while at the same time acting on the ion's motion. In effect, the laser induced radiation pressure leads, for the appropriate detuning, to a damping of the motion. 

\subsection{\textbf{Excitation source characterisation}}
\label{sec:excitation_source}
The axial motion of the ion is excited by applying a square pulse train with variable amplitude, duration and repetition rate to one of the dc-electrodes. In order to maximise the frequency resolution we have investigated these three parameters.

\subsubsection{\textbf{Excitation pulse amplitude}}
\label{sec:ExcitationAmplitude}

\begin{figure}
\begin{center}
\includegraphics[width=0.6\linewidth]{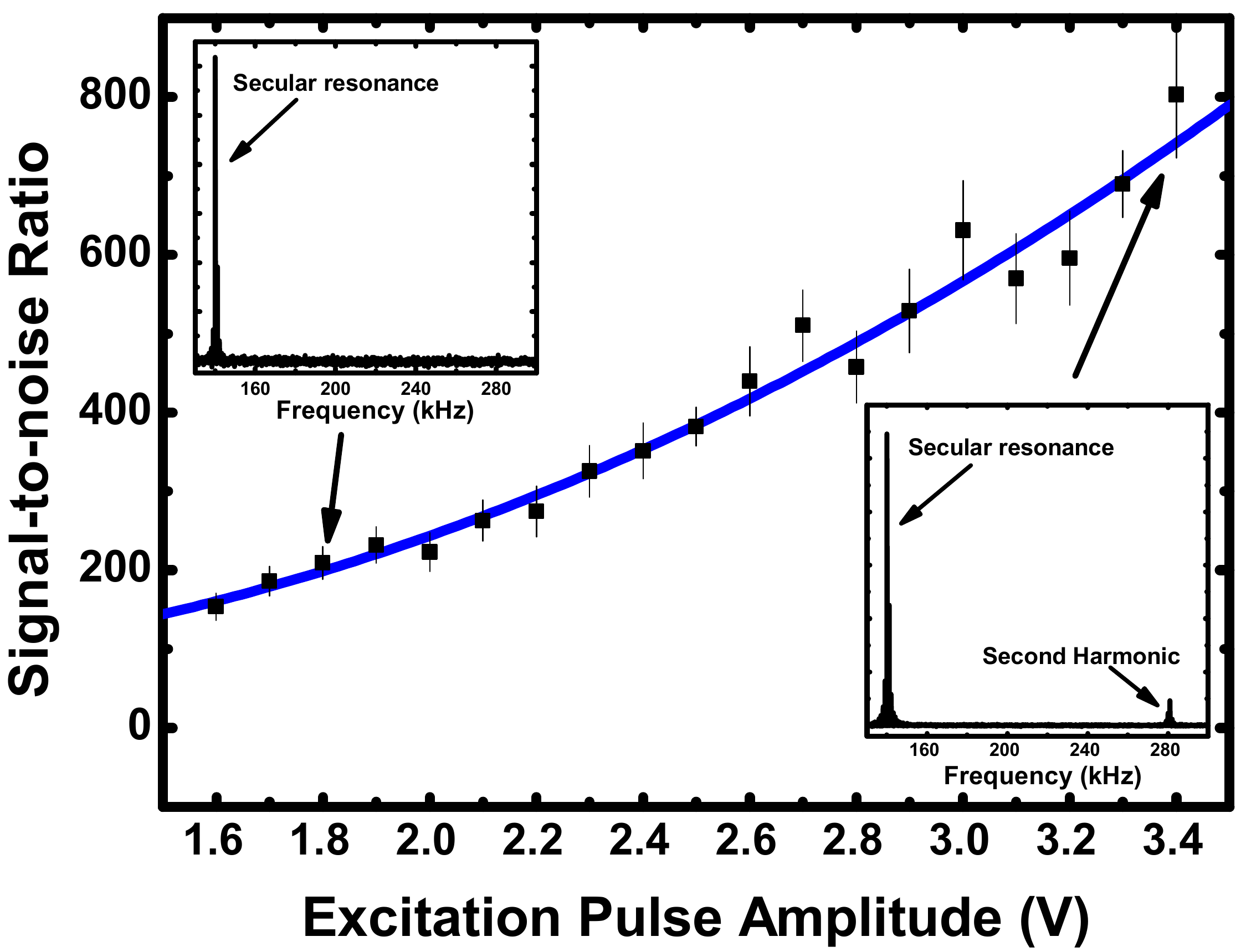}
\caption{Amplitude of the Lorentzian fit of the auto-correlation spectrum as a function of excitation pulse amplitude for a single ion. The interrogation time for each measurement is 20~s, the repetition rate is 1~kHz, the pulse width is 3~$\mu$s, the laser detuning is 30~MHz and the laser power is 10~$\mu$W. The insets shows the auto-correlation spectra for excitation pulse amplitudes of 1.8~V and 3.4~V.}
\label{fig:AmplVsPulseAmpl}
\end{center}
\end{figure}

For short excitation pulses, the laser induced damping during the pulse can be neglected, which leads to a linear increase of the ion's elongation in the trapping potential with increasing excitation pulse amplitude. This in turn results in an oscillation amplitude which is proportional to the excitation amplitude. According to equation \ref{equ:fluoresMod1}, the ion's oscillation is converted into a modulation of its fluorescence. The auto-correlation, however, is a bilinear function of the modulation. Thus, the peak height of the motional spectrum, obtained from the Fourier transform of the auto-correlation, increases quadratically with the excitation amplitude. This is clearly visible in the measurements shown in figure \ref{fig:AmplVsPulseAmpl} together with a parabolic fit. 
For large excitation amplitudes equation \ref{equ:fluoresMod1} is not valid and the conversion between the ion's oscillation and its fluorescence becomes non-linear. This leads to the occurrence of harmonics of the secular frequency in the auto-correlation spectrum. The measured spectra in figure \ref{fig:AmplVsPulseAmpl} (inset) show that for small amplitudes no harmonic of the secular frequency is visible, however, a second peak at twice the frequency occurs for larger amplitudes.
 
Even though a large excitation amplitude leads to favorable signal-to-noise ratios it has an adverse effect on the ion's temperature which may limit the excitation amplitudes for some applications.

\subsubsection{\textbf{Excitation pulse width}}
\label{sec:ExcitationPulseWidth}

\begin{figure}
\begin{center}
\includegraphics[width = 0.6 \linewidth]{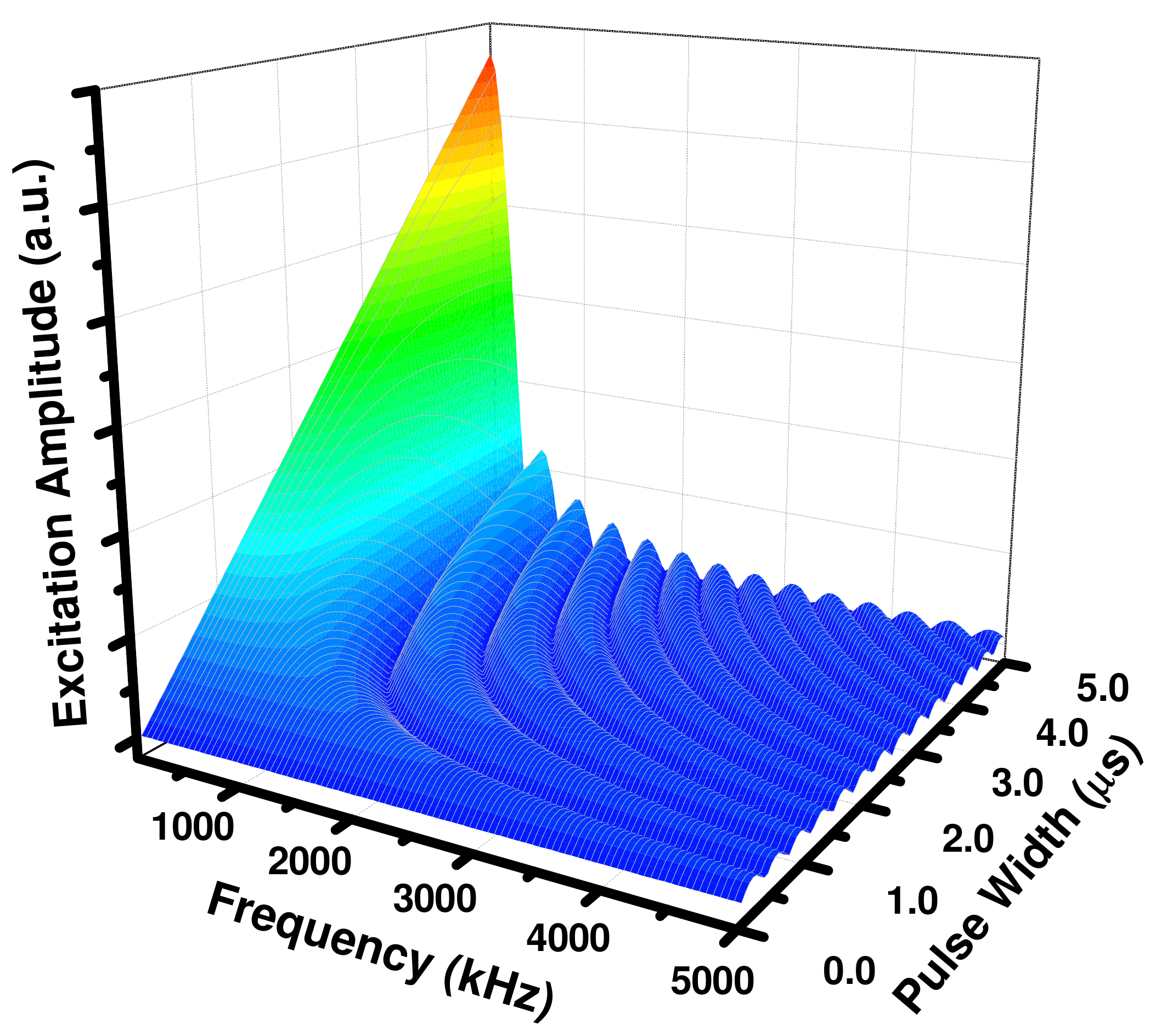}
\caption{Excitation amplitude as a function of secular frequency and excitation pulse width.}  
\label{fig:pulse_width_sim}
\end{center}
\end{figure}

\begin{figure}
\begin{center}
\includegraphics[width = 0.6 \linewidth]{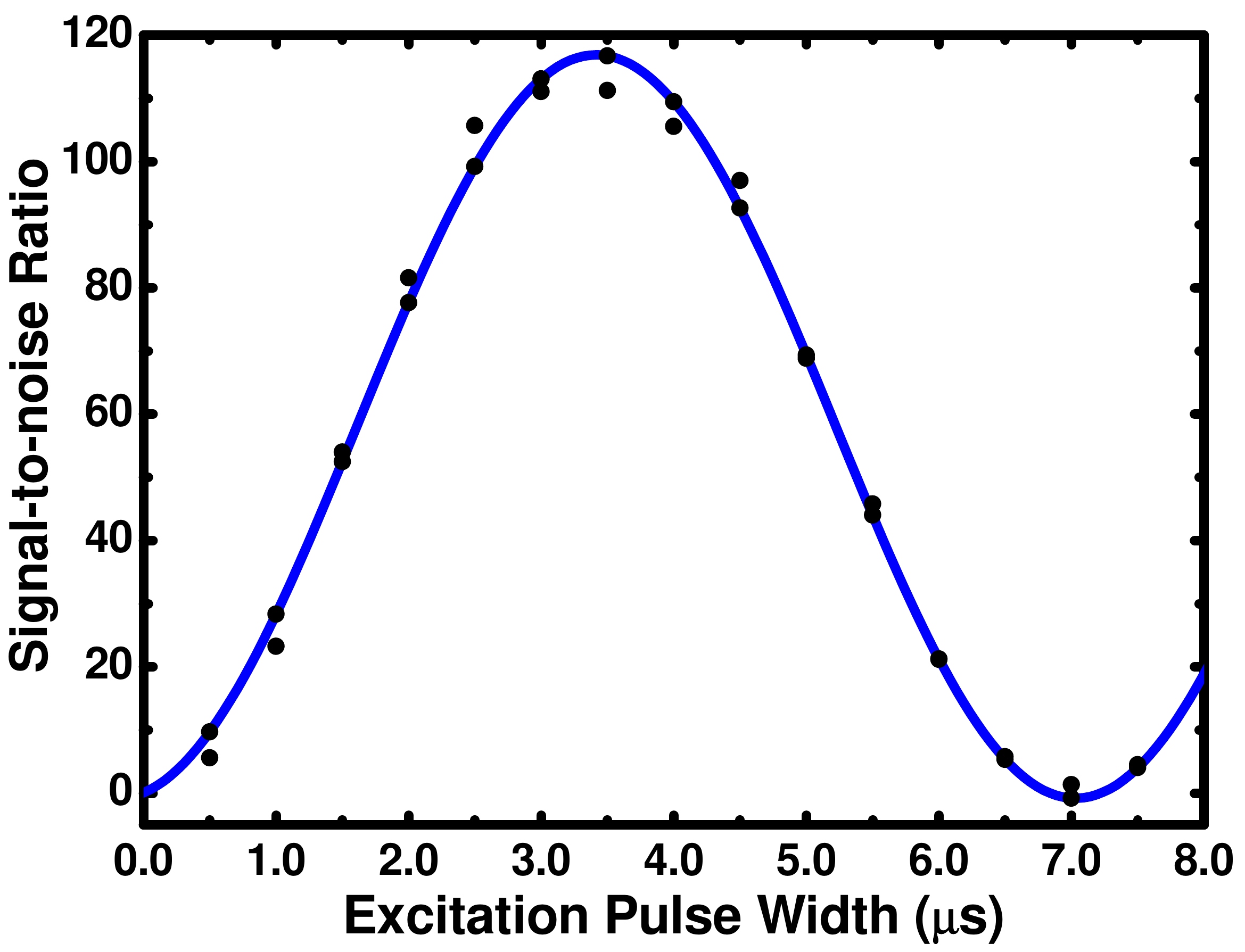}
\caption{SNR of the auto-correlation spectra as a function of excitation pulse width for a single ion. The interrogation time is 10~s, the repetition rate is 1~kHz and the excitation amplitude is 2~V. A sine squared function has been fitted to the data.}  
\label{fig:mod_pulse_width}
\end{center}
\end{figure}

The periodic excitation has a spectrum consisting of a comb-like structure with periodic peaks separated by the repetition rate as described by equation \ref{equ:combspectrum}. Its excitation amplitude spectrum is shown in Figure \ref{fig:pulse_width_sim} for a range of frequencies and pulse widths. Only frequencies that do not coincide with the zero crossings at $\omega_{sec}\tau=2\pi$ can be detected and the excitation amplitude inevitably decreases with frequency. 
For a fixed secular frequency of the ion, the excitation amplitude exhibits a sinusoidal oscillation as $\sin\!\left(\omega_{sec}\tau/2\right)$ (see equation \ref{equ:combspectrum}). We have measured this behaviour for an axial secular frequency of 140~kHz for a range of modulation pulse widths from 0.5~$\mu$s up to 7.5~$\mu$s. Figure \ref{fig:mod_pulse_width} shows the SNR of auto-correlation spectra for the measured range of excitation pulse widths. Due to the quadratic dependency of the signal on the excitation amplitude (see section \ref{sec:ExcitationAmplitude}), the height of the auto-correlation spectrum changes as $\sin^2\!\left(\omega_{sec}\tau/2\right)$. 
The data agree well with equation \ref{equ:combspectrum} which predicts a maximum at a pulse width of 3.57~$\mu$s and a minimum at 7.1~$\mu$s.

For a given secular frequency, the optimum excitation depends on both the pulse width and the excitation amplitude. Since the secular frequency may not be well known or may change with time, long pulse widths are undesirable as they lead to a narrow excitation spectrum. Employing short pulses increases the spectral width but leads to a decrease of the excitation amplitude. In order to compensate this, the pulse amplitude must be increased which may have an adverse effect on the temperature of ion crystals. Even though the excitation couples mainly to the COM-mode, coupling to higher modes of ion crystals or strings cannot be fully suppressed. Careful balancing of the pulse parameters is required to obtain a wide excitation spectrum while at the same time maintaining a low temperature of the trapped ions during the measurement.

\subsubsection{\textbf{Excitation pulse repetition rate}}
\label{sec:rep_rate}

\begin{figure}[h]
\begin{center}
\includegraphics[width = 1 \linewidth]{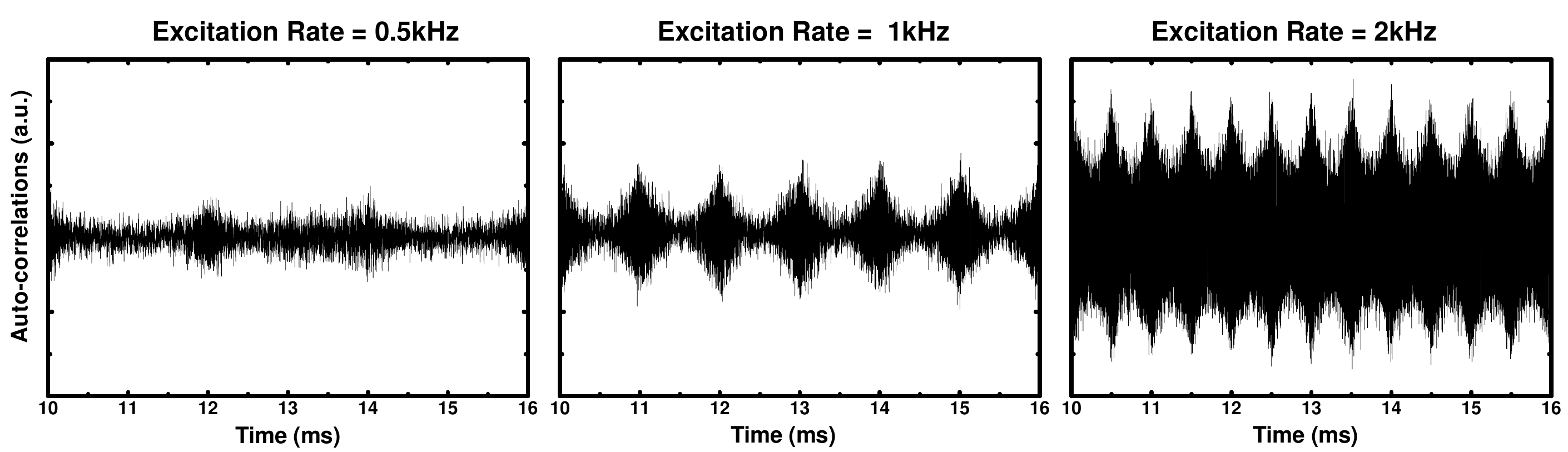}
\caption[]{Auto-correlation measurements for three repletion rates. In all measurements the laser power is 10~$\mu$W, the detuning is 30~MHz, the excitation amplitude is 2~V and the pulse width is 3~$\mu$s. } 
\label{fig:mod_pulse_rep}
\end{center}
\end{figure}

Increasing the repetition rate of the excitation has two effects on the auto-correlation spectra. Firstly, the spacing between the comb peaks in the spectrum becomes larger since the separation reflects the repetition rate. Thus, for an accurate determination of the motional frequency of the ion small repetition rates seem advantageous as the spectral resolution increases. 
The second effect comes from the impact of the repetition rate on the signal amplitude. Each voltage pulse excites the oscillation of the trapped ion which is damped by the interaction with the laser light. Following each pulse the oscillation amplitude of the ion, and therefore the signal, decays. If the decay rate is much faster than the repetition rate of the excitation, the average of the signal over one excitation period is small, which in turn results in a small peak height of the auto-correlation spectrum. Due to the quadratic dependence of the height of the auto-correlation spectrum on the ion's oscillation amplitude, the signal is proportional to the square of the average amplitude which is described by the first term in equation \ref{equ:reprate}. 
In addition, the number of correlation events contributing to the signal increases linearly with the ratio of the interrogation time to pulse period and thus with the repetition rate. Both effects are clearly visible in figure \ref{fig:mod_pulse_rep} in which auto-correlation measurements for three different repetition rates are shown. 
While for a repetition rate of 500~Hz the signal completely decays before another excitation pulse is applied, the oscillation overlaps significantly with the next pulse for a rate of 2~kHz. The increase of the overall amplitude is due to the increase in correlation events for larger repetition rates.

\begin{figure}
\begin{center}
\includegraphics[width = 0.6 \linewidth]{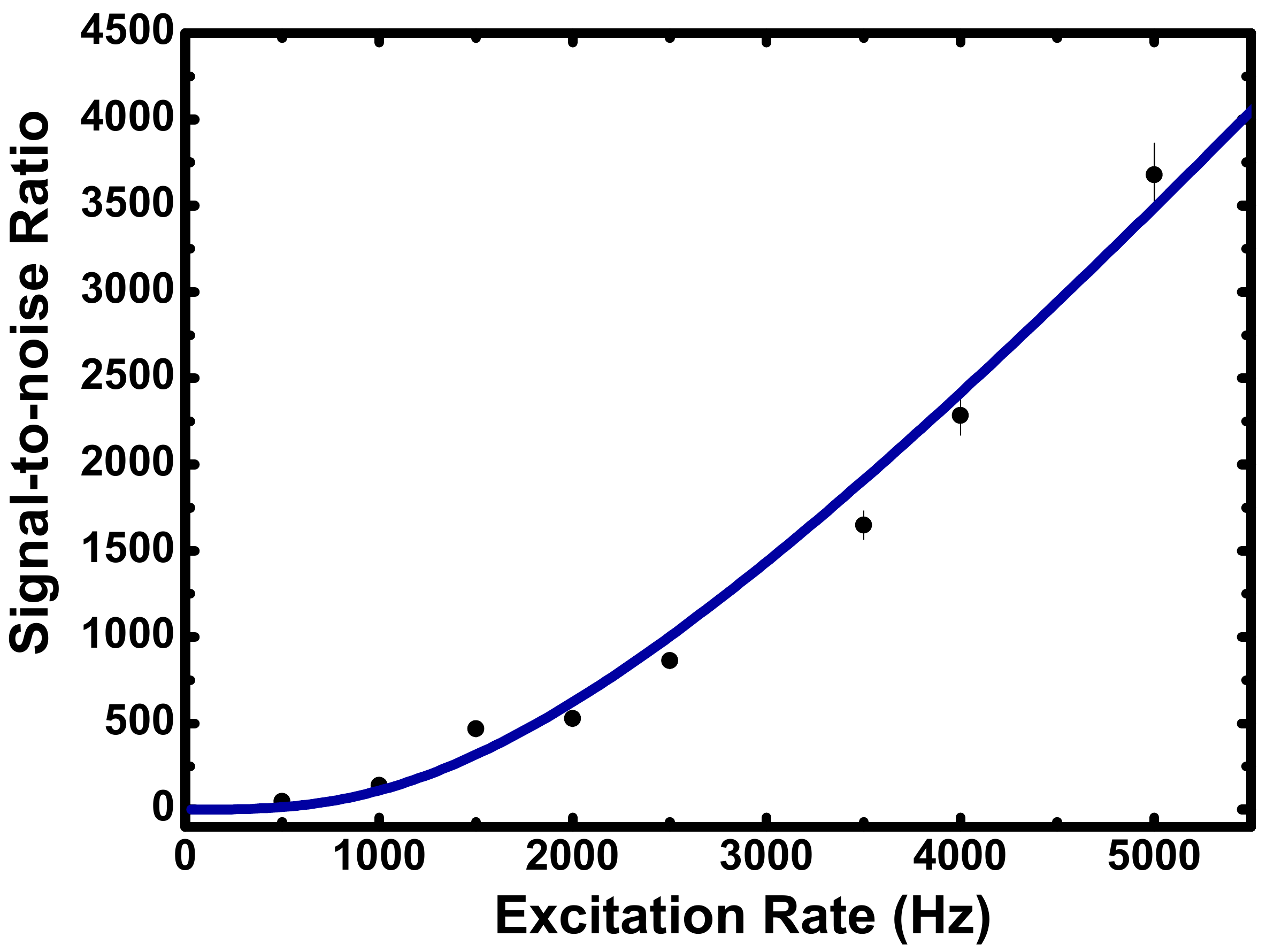}
\caption{SNR of the auto-correlation spectrum for a range of excitation pulse repetition rates measured with a single ion. The laser detuning is 30~MHz, the laser power is 10~$\mu$W, the excitation amplitude is 2~V and the pulse width is 3~$\mu$s.} 
\label{fig:mod_pulse_repD}
\end{center}
\end{figure}

In the regime where the damping is linear, e.g. for small oscillation amplitudes, the dependence of the auto-correlation signal amplitude on the repetition rate $R$ can be approximated as:
\begin{equation}
\mbox{Signal}\propto \underbrace{\left\{\frac{R}{\gamma}\left(1-\exp\left(-\frac{\gamma}{R}\right)\right)\right\}^2}_{\mbox{\tiny average signal}}\cdot\underbrace{R}_{\mbox{\tiny correlation events}}
\label{equ:reprate}
\end{equation}
where $\gamma$ is the damping rate of the ion's motion. Figure \ref{fig:mod_pulse_repD} displays the measured signal amplitudes of the auto-correlation spectrum for a range of repetition rates and a fit of equation \ref{equ:reprate} to the data. 
Even though the signal increases with increasing repetition rate, the accuracy of measuring the secular frequency is at the same time adversely affected by the increased spacing of the comb peaks in the excitation spectrum.
 
The effect of the repetition rate on the measurement accuracy depends on the way the spectrum is analysed. The easiest way of analysing the data is by using the highest comb peak in the auto-correlation spectrum as the secular frequency. In this case the error is half the repetition rate and an increase in repetition rate directly degrades the spectral resolution. This method does enable very fast measurements since a reliable signal can be obtained for a SNR as low as 1. For example, we have achieved a SNR of 12 for a 200-ion crystal with an interrogation time of only 102~ms using a repetition rate of 500~Hz. By fitting a Lorentzian to the comb peaks the accuracy can be significantly higher than the repetition rate. A requirement for fitting a Lorentzian is that there are at least three visible comb peaks in the spectrum. For repetition rates smaller than the width of the auto-correlation resonance, many comb peaks are available increasing the accuracy to which the motional resonance can be measured. In contrast, for repetition rates much higher than the motional damping rate, there is mainly one comb peak of the excitation spectrum contributing to the signal with the other peaks being quadratically suppressed with increasing repetition rate. 

A detailed analysis shows that the error in the frequency measurement is proportional to $\frac{R}{H^2}$, with the centre peak height of the auto-correlation spectrum $H$.
Taking equation \ref{equ:reprate} into account, the frequency error decreases with increasing repetition rate. The effect of the enhanced signal for large repetition rates offsets the reduction of the comb peaks due to their increased detuning from the auto-correlation resonance. However, this is only valid as long as $R\ll\omega_{sec}$.

\subsection{\textbf{Laser parameters}}
\label{sec:laser_parameters}

While the pulsed excitation acts only on the ion's motion, the laser also has an effect on the motional states of the ion. 
The laser power and detuning affect the conversion of the ion's motion into a modulation of its fluorescence by influencing the ion's state population and thus $\left.\frac{d\rho_{ee}}{d\delta}\right|_\delta$ (see equation \ref{equ:fluoresMod1}).
However, the velocity dependent radiation pressure directly affects the motion of the ion. For red detuned lasers close to resonance this results in a damping of the ion's oscillation in the trap and thus gives rise to a finite line width of the motional spectrum. For small velocities the damping rate of the ion's oscillation is proportional to $\left.\frac{d\rho_{ee}}{d\delta}\right|_\delta$. Therefore, the laser induced damping interferes with the detection of the ion's motion.

\subsubsection{\textbf{Laser detuning}}
\label{sec:detuning}

\begin{figure}[h]
\begin{center}
\includegraphics[width = 0.8 \linewidth]{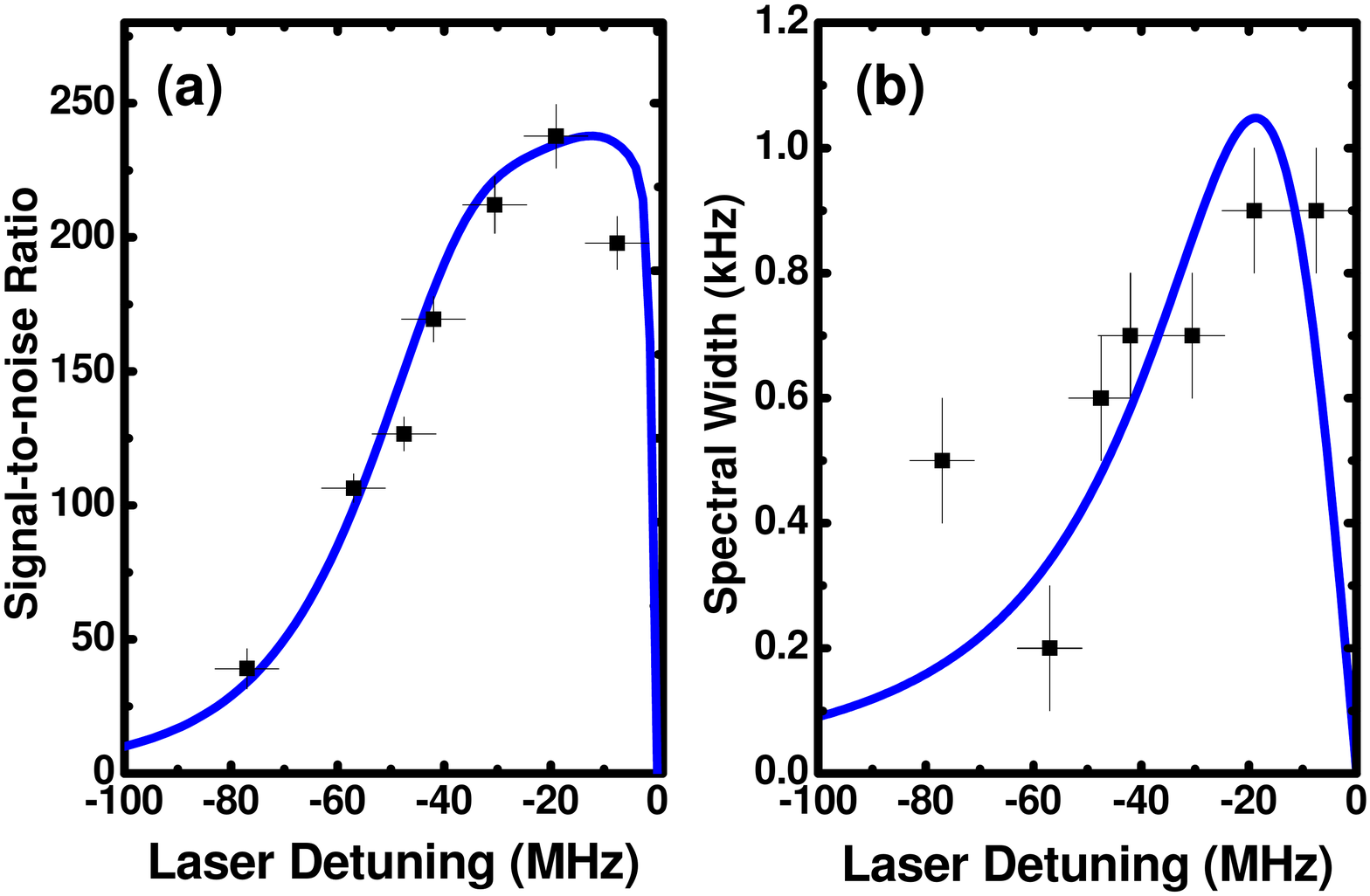}
\caption{SNR and width of the auto-correlation spectrum as a function of the cooling laser detuning measured with a single ion. The laser power is 5~$\mu$W, the excitation amplitude is 1.6~V, the pulse width is 3~$\mu$s and the repetition rate is 500~Hz.} 
\label{fig:laser_detuning_ampl}
\end{center}
\end{figure}

According to equation \ref{equ:fluoresMod1} the optimal conversion efficiency between the ion's motion and its fluorescence is obtained by tuning the laser to the largest gradient of the ion's spectral line profile. However, at this detuning, the laser induced damping has a simultaneous maximum which in turn leads to a reduction of the oscillation amplitude and thus to a reduction in the height of the auto-correlation spectrum. By tuning the laser closer to resonance, the decrease of the conversion efficiency is offset by a reduction of the damping leading to a shift of the optimal detuning towards the atomic resonance. The dependence of the signal strength on the detuning is shown in figure \ref{fig:laser_detuning_ampl}(a). The laser induced damping shifts the optimal detuning towards the atomic resonance and leads to a plateau at small detuning. This can be seen by comparing the position of the maximum SNR with the maximum in the width of the auto-correlation spectrum (see figure \ref{fig:laser_detuning_ampl}(b)). The width of the spectrum corresponds to the damping of the ion's motion and thus the slope of its spectral line.  This is illustrated in figure \ref{fig:laser_detuning_ampl} were the maximal width is further detuned from resonance compared to the optimal SNR.
The measurements shown in figure \ref{fig:laser_detuning_ampl} are fitted with a full model of the system. The SNR data show very good agreement with the model and the spectral width measurements show a reasonable agreement. Since the spectrums are narrow (less than 1~kHz), there is increased error in fitting a Lorentzian envelope to the limited number of discrete peaks forming each spectrum.

\subsubsection{\textbf{Laser power}}
\label{sec:laser_power}

\begin{figure}[h]
\begin{center}
\includegraphics[width = 0.8 \linewidth]{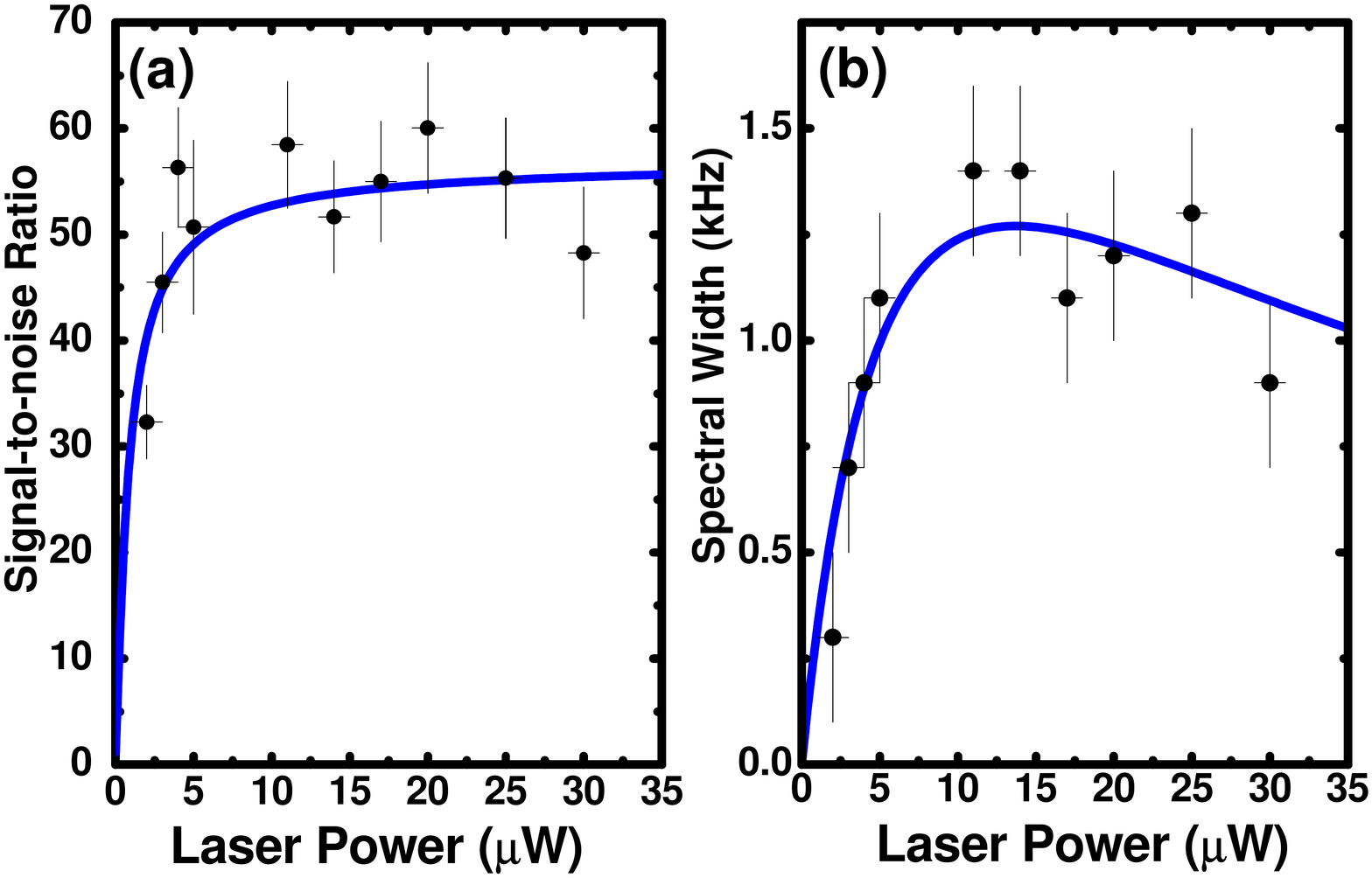}
\caption{SNR and width of the auto-correlation spectrum as a function of the cooling laser power measured with a single ion. The excitation pulse amplitude is 2~V, the repetition rate is 500~Hz, the pulse width is 3~$\mu$s and the laser detuning is 40~MHz.}
\label{fig:laser_power_ampl}
\end{center}
\end{figure}

The effect of the laser induced damping of the ion's motion can also be observed in the dependence of the SNR and the width of the auto-correlation spectrum on the laser power which is shown in Figure \ref{fig:laser_power_ampl}.
For small laser power the signal increases fast and reaches a plateau as illustrated by the data in figure \ref{fig:laser_power_ampl}(a). In contrast, the spectral width increases for small laser intensities due to the increased damping and after reaching a maximum it decreases steadily. 
For small laser intensities, the slope of the ion's spectral line is governed by the fluorescence level while the width of the transition corresponds to the natural line width. Thus, with increasing intensity the signal and fluorescence level increase together. For large laser intensities, however, the fluorescence level saturates while the transition line width is governed by saturation broadening. This results in a decrease of the gradient of the spectrum and thus a decrease in the damping.
This reduced damping, in turn, increases the ion's oscillation amplitude. Hence, the height of the auto-correlation spectrum is increased and this compensates the decrease in the conversion efficiency. 

The motional spectrum SNR also relies upon the fluorescence signal to background scatter ratio. For a laser detuning of 20~MHz and a power of 15~$\mu$W, we measure a fluorescence signal to background scatter ratio of 5 for a single ion.

\subsection{\textbf{Heating of the ion}}
\label{sec:heating}

For many applications it is critical that the heating of the ion during the measurement is very small. This is especially important if the secular frequency of an entire ion crystal is to be detected. Typically, trapped ions form crystals at temperatures in the order of 100~mK \cite{Drewsen2}. Thus, the average kinetic energy imparted to the crystal by the measurement must be lower than this. For small oscillation amplitudes, the amplitude of the fluorescence modulation can be approximated by equation \ref{equ:fluoresMod2} which we use to determine the kinetic energy of the ion. By taking the ratio of the auto-correlation's modulation amplitude and its constant offset, the detection efficiency cancels and the oscillation amplitude can be determined by using the spectroscopic properties of the transition which have been independently measured. The maximal energy induced into the COM-mode is calculated from the maximal amplitude of the oscillation whereas the average energy is determined by taking the decay of the oscillation into account.

\begin{figure}[h]
\begin{center}
\includegraphics[width = 0.6 \linewidth]{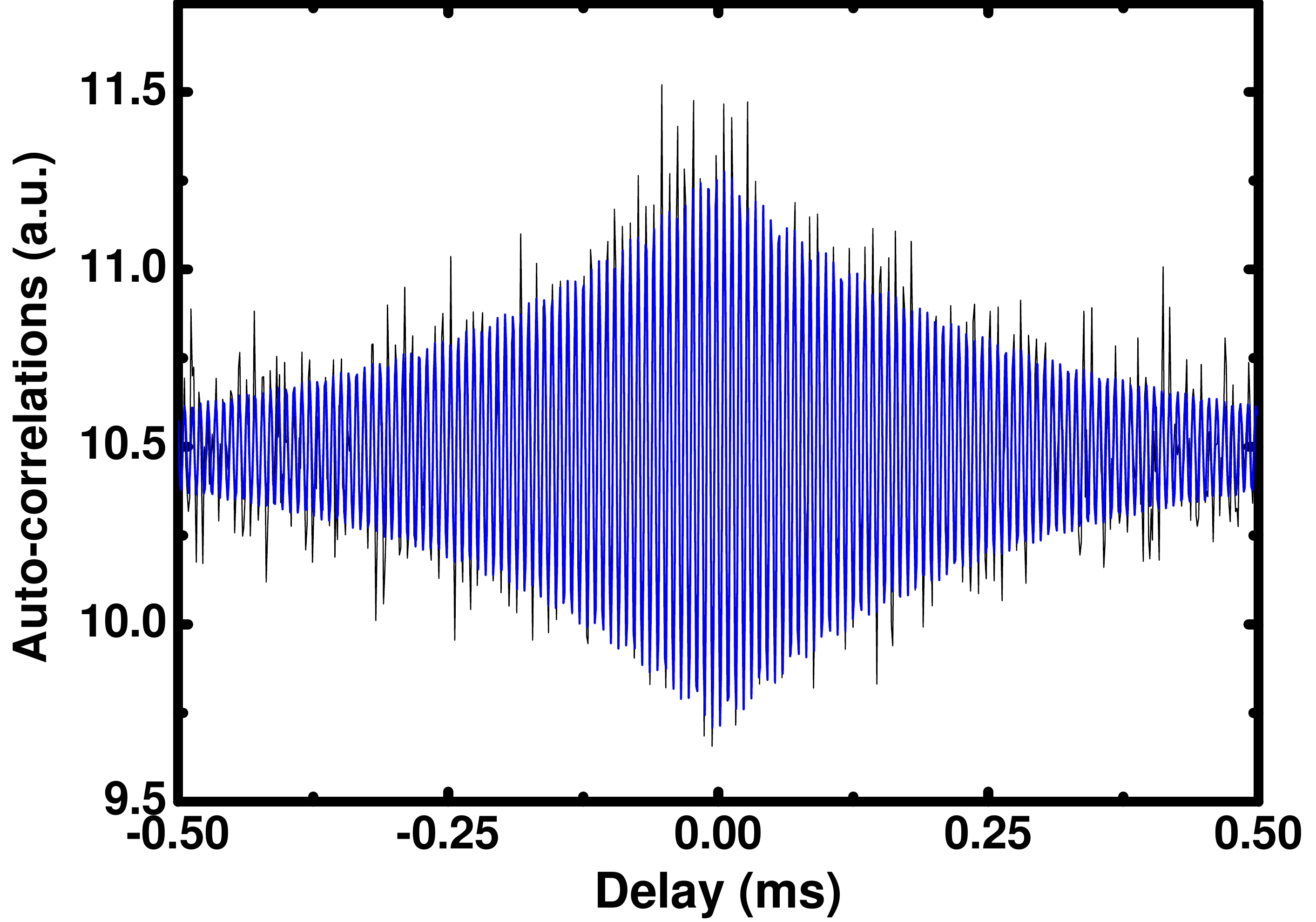}
\caption{Auto-correlation measurement including a fitted damped oscillation. The laser detuning is 30~MHz, the laser power is 5~$\mu$W, the excitation amplitude is 2~V, the repetition rate is 500~Hz and the pulse width is 3~$\mu$s.} 
\label{fig:auto-correlation2}
\end{center}
\end{figure}

Figure \ref{fig:auto-correlation2} shows an auto-correlation measurement used to determine the measurement induced kinetic energy of a single ion. To obtain the oscillation amplitude, a damped oscillation has been fitted to the data. 
From the analysis of this measurement we have obtained a maximal kinetic energy of 487(22)~neV introduced into the ion's motion. This corresponds to a temperature increase of 11.3(5)~mK. The average measurement induced kinetic energy of the ion is around 150~neV corresponding to a temperature increase of about 3~mK. This is low enough for most applications and enables the use of this technique with larger ion crystals. However, the heating depends on the excitation parameters which must be carefully chosen in order to obtain a good signal but at the same time keep the ion's temperature low.

\subsection{\textbf{Summary of measurement characterisation}}
\label{sec:characterisation_summary}
 
We have characterised all of the system parameters and shown how each contributes to the SNR of the auto-correlation spectrum. The optimum laser parameters are clearly illustrated by the measurements shown in figure \ref{fig:laser_detuning_ampl} and figure \ref{fig:laser_power_ampl}. The optimum excitation pulse parameters, however, will depend upon the details of the desired application, e.g.\ the frequency range being measured and the limit placed on the energy imparted to the system by the interrogation. We have shown that the careful balancing of these parameters provides a high contrast motional spectrum (SNR$>$100) of a single ion with a low average temperature increase (around 3~mK) in a short interrogation time (less than 30~s).

\section{Application to linear ion strings and three dimensional crystals}
\label{sec:applications}

In the following section we present applications of the above described secular frequency measurement technique. We show how the technique can be used to identify the mass of sympathetically cooled dark ions and measure the configuration shifts of a mixed linear ion string. In addition, we apply the technique to a three-dimensional crystal and use it to continuously monitor the mass of the crystal to observe charge exchanges between calcium isotopes.

\subsection{\textbf{Ion mass identification and configuration measurements}}
\label{sec:ion_mass_identification_and_configs}

By comparing the COM-mode frequency of a mixed species ion string to the frequency of an ion with known mass it is possible to identify the mass of the dark ion contained within the string. Furthermore, we can resolve the shifts in the axial secular frequency resulting from the different configurations of a mixed species ion string.

\subsubsection{\textbf{Ion mass identification}}
\label{sec:ion_mass_identification}

Due to the dependence of the secular motion on the charge-to-mass ratio of an ion, the measurement of its motional frequency can be used to identify the trapped ion species \cite{Drewsen} if the charge of the ion is known. To measure the mass of a single dark ion, we trap it alongside a $^{40}$Ca-ion which is used to sympathetically cool the ion and to detect its secular frequency. The COM-mode frequency of this two-ion string can be expressed by 
\begin{equation}
\nu_{\mbox{\tiny{COM}}}^{2}=\left[(1+M_{0}/M_{1})-\sqrt{1-M_{0}/M_{1}+\left(M_{0}/M_{1}\right)^2}\right]\nu^{2}_{0},
\label{equ:string_freq}
\end{equation}
if the charge of the two ions is the same \cite{Morigi}. Thus, if the secular frequency $\nu_{0}$ of an ion with known mass $M_{0}$ is measured, the mass of the dark ion $M_{1}$ can be determined by measuring the ion string's COM-mode frequency.

We have measured the COM-mode frequency shift with respect to a single $^{40}$Ca-ion for a two-ion string in which a $^{40}$Ca-ion is trapped alongside other calcium isotopes. For each isotope, a series of measurements has been performed with a driving pulse amplitude of 3.5~V, repetition rate of 500~Hz, pulse width of 3~$\mu$s and interrogation time of 30~s. The measured frequencies and the resulting masses of $^{43}$Ca and $^{44}$Ca are listed in table \ref{tab:isotope_frequencies} together with the literature values.
\begin{table}[h]
\begin{center}
\begin{tabular}{| c | c | c |}
\hline
 Isotope & Measured Mass (amu)& Literature Value (amu) \cite{Emsley}\\  \hline
$^{43}$Ca$^{+}$ &  42.96(7) & 42.9587662(13) \\ \hline
$^{44}$Ca$^{+}$ &  43.90(8) & 43.9554806(14) \\ \hline
\end{tabular}
\caption{The measured masses of two calcium isotopes and their literature values. }
\label{tab:isotope_frequencies}
\end{center}
\end{table}

The results demonstrate that the mass can be measured a precision of better than $2\cdot 10^{-3}$. This precision is only limited by a slow drift of the secular frequency due to the temperature change of the trap during the loading process.

\subsubsection{\textbf{Measurement of configuration shifts in a linear ion string}}
\label{sec:configurations}

Since the axial frequency measurement technique we employ induces only little heating into the ion configuration, it is possible to measure the COM-mode frequencies of individual configurations of a mixed species linear ion string. The perturbation of the linear ion string due to the presence of an ion with different mass leads to a shift of the COM-mode frequency which depends not only on the average charge-to-mass ratio of the ions but also on the position of the ions in the string \cite{Morigi}. 

A measurement of the configuration shifts in a mixed linear ion string is performed for a five ion string consisting of four $^{40}$Ca-ions and one dark ion. The dark ion used has a mass substantially larger than the mass of a $^{40}$Ca-ion. In order to distinguish the five possible configurations of the ion string, it is monitored with a CCD camera. 

\begin{figure}[ht]
\begin{center}
\includegraphics[width = 0.7 \linewidth]{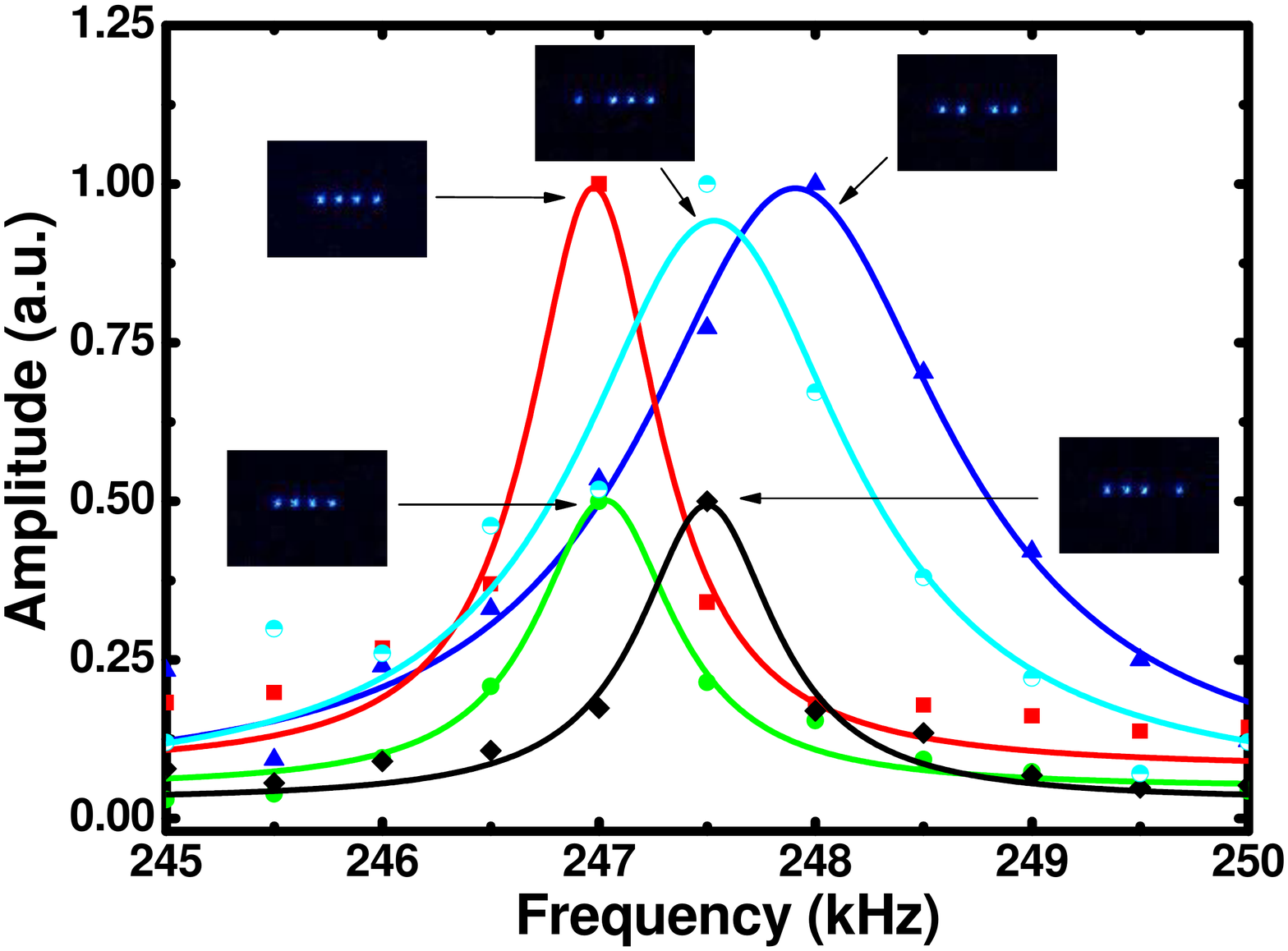}
\caption{Normalised motional spectra measured for each of the five possible configurations of the mixed ion string. Two spectra, given by the symmetric configurations, have been rescaled for clarity. A Lorentzian function is fit to each spectrum. The measurement interrogation time is 15~s, with an excitation pulse amplitude of 1.5~V, a repetition rate of 500~Hz and a pulse width of 3~$\mu$s.}
\label{fig:configs_experiment}
\end{center}
\end{figure}

The COM-mode frequency has been measured for each of the five configurations several times and the average of the measurements has been taken. The axial COM-mode frequency and the standard deviation for each configuration is listed in table \ref{tab:configs_frequencies}. There are three configurations giving a unique axial frequency shift which are labeled A, B and C. Configurations D and E are equivalent to configurations A and B, and give the same measured axial frequency shift accordingly. Configuration F is the pure string consisting of five $^{40}$Ca-ions. Figure \ref{fig:configs_experiment} shows one motional spectrum for each configuration together with a Lorentzian fit.
\begin{table}[h]
\begin{center}
\begin{tabular}{| c | c | c |}
\hline
 & Configuration & Axial Frequency (kHz) \\  \hline
A& \includegraphics[scale=0.1]{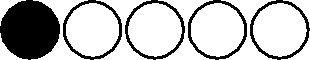}& 247.08(17) \\ \hline
B& \includegraphics[scale=0.1]{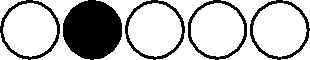} & 247.47(18) \\ \hline
C& \includegraphics[scale=0.1]{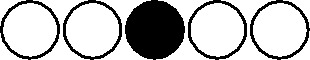}& 247.81(16) \\ \hline
D& \includegraphics[scale=0.1]{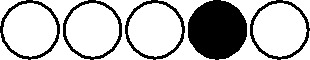}& 247.50(17) \\ \hline
E& \includegraphics[scale=0.1]{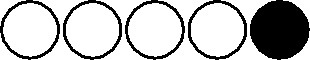}& 247.11(19) \\ \hline
F& \includegraphics[scale=0.1]{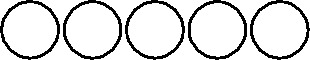}&  258.46(3) \\ \hline
\end{tabular}
\caption[]{The experimentally measured axial frequencies for each of the five possible configurations of an ion string consisting of four $^{40}$Ca-ions and one dark ion and a pure crystal containing five $^{40}$Ca-ions.
\label{tab:configs_frequencies}}
\end{center}
\end{table}

\begin{figure}
\begin{center}
\includegraphics[width = 0.6 \linewidth]{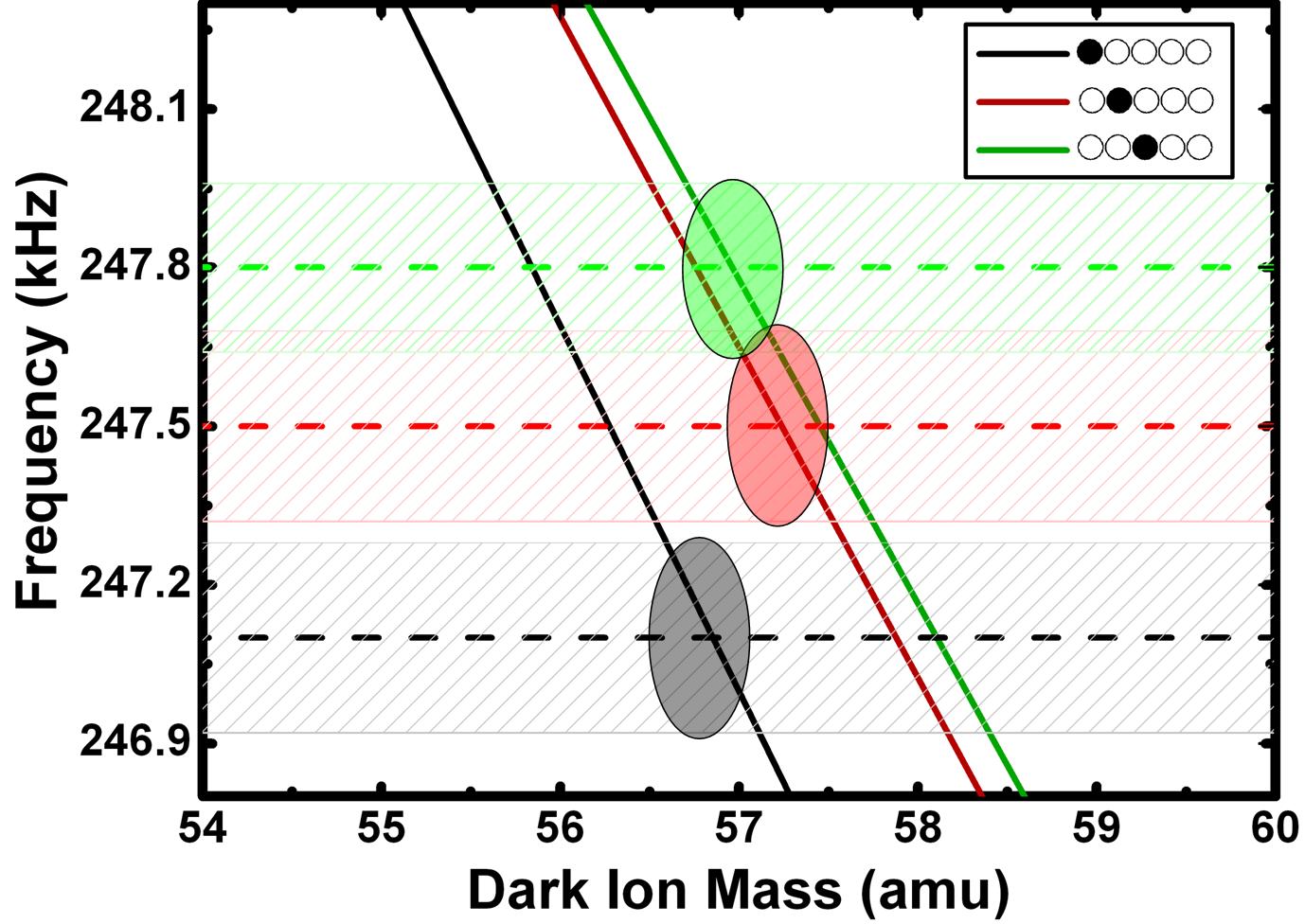}
\caption{Simulation (solid lines) comparing the axial frequency for the three unique configurations in an ion string containing four $^{40}$Ca-ions and one dark ion as a function of the molecular weight of the dark ion. Experimentally measured frequencies (dotted horizontal lines) are included with bands indicating the measurement error.} 
\label{fig:configs_simulation}
\end{center}
\end{figure}

A simulation is performed to calculate the theoretical shift of the axial frequency from the experimentally measured frequency of the pure $^{40}$Ca-ion string for each of the configurations as a function of the mass of the dark ion. The result of the simulation is shown in figure \ref{fig:configs_simulation} together with the measured frequencies (dotted horizontal lines.) The bands around the measured frequencies indicate the measurement errors and the ellipses show the overlap with the simulations.

Table \ref{tab:configs_dark_mass} lists the dark ion masses which correspond to the experimentally measured frequency shift for the three unique configurations. Averaging these three masses gives 57.0(3)~amu. This is within experimental error of the calcium monohydroxide molecular ion, which has a molecular weight of 57.085~amu. 

\begin{table}[h]
\begin{center}
\begin{tabular}{| c | c |}
\hline
 Configuration & Dark Ion Mass (amu) \\  \hline
\includegraphics[scale=0.1]{config_A}& 56.8(2) \\ \hline
\includegraphics[scale=0.1]{config_B} & 57.2(3) \\ \hline
\includegraphics[scale=0.1]{config_C}& 56.9(3) \\ \hline
\end{tabular}
\caption[]{The dark ion mass values determined from comparing the experimental configuration shift to the shifts given by the simulation, shown in figure \ref{fig:configs_experiment}.
\label{tab:configs_dark_mass}}
\end{center}
\end{table}

\subsection{\textbf{Measurement of the COM-mode frequency of a large crystal}}
\label{sec:COM_mode_large_crystal}

\begin{figure}[h]
\begin{center}
\includegraphics[width = 0.8\linewidth]{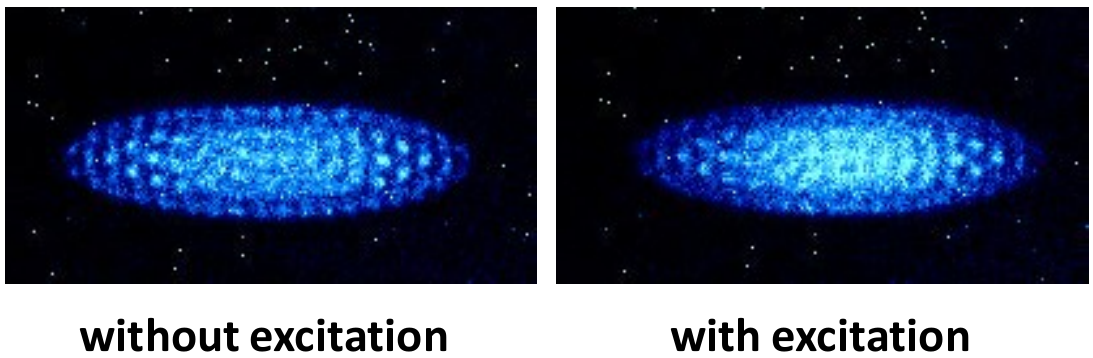}
\caption{A large three dimensional Coulomb crystal made up of a few hundred$\;^{40}$Ca ions.}
\label{fig:large_crystal_pic_pdf}
\end{center}
\end{figure}

\begin{figure}[h]
\begin{center}
\includegraphics[width = 0.8 \linewidth]{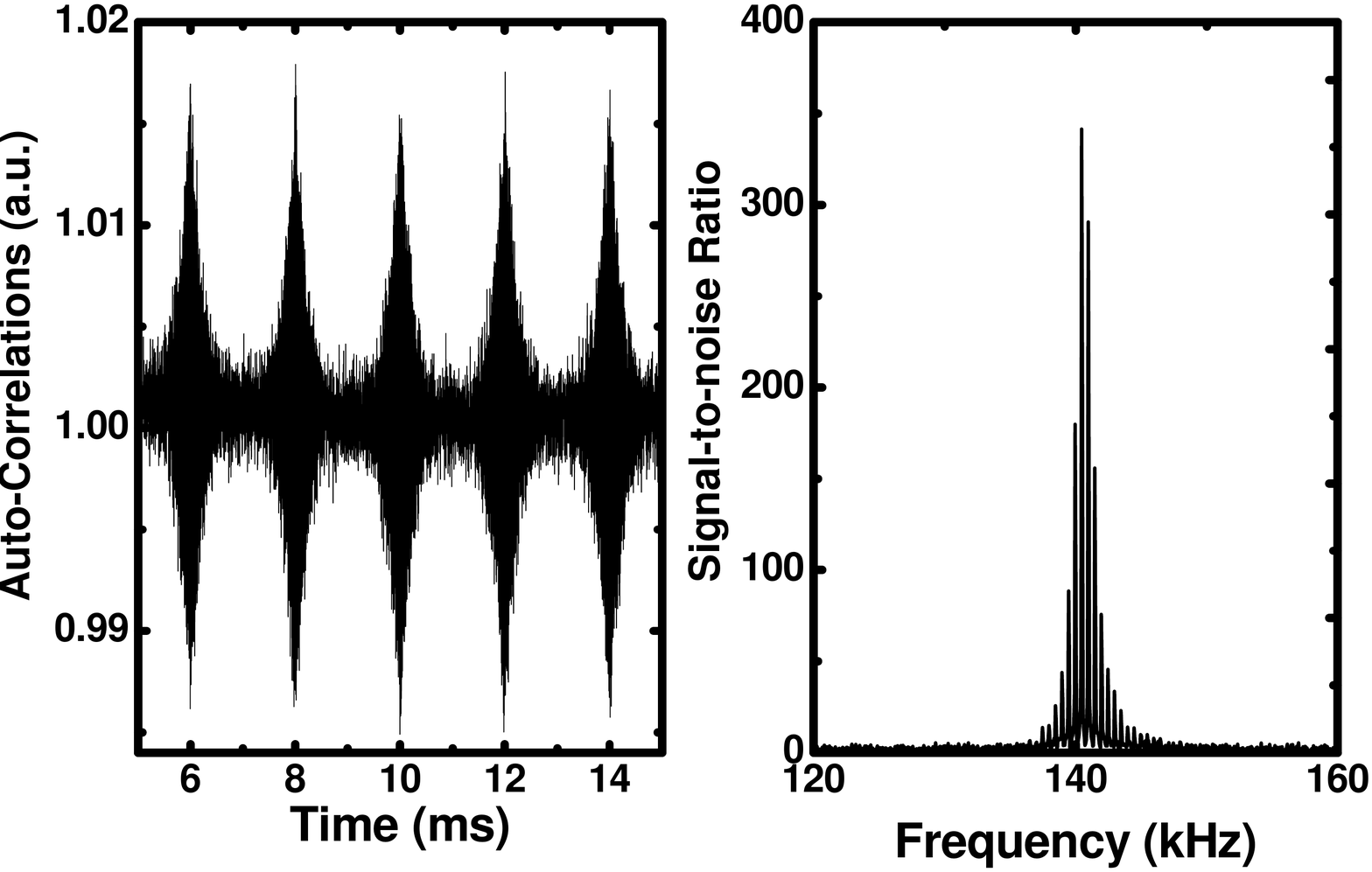}
\caption{Auto-correlation measurement and motional spectrum of a three-dimensional ion crystal. The laser detuning is 150~MHz, the laser power is 25~$\mu$W, the excitation amplitude is 2~V, the repetiton rate is 500~Hz, the pulse width is 3~$\mu$s and the interrogation time is 5~s.}
\label{fig:crystal_g2_and_spec}
\end{center}
\end{figure}

The secular frequency measurement described above for single ions and short ion strings can also be applied to large three-dimensional Coulomb crystals. Using the same driving pulse parameters, the structure of the crystal remains completely intact during the measurement. Figure \ref{fig:large_crystal_pic_pdf} shows a three-dimensional Coulomb crystal with and without excitation. The slight blurring of the image of the crystal with modulation indicates the motion of the ions. However, the crystal structure is still intact and the COM-mode energy corresponds to a measurement induced average temperature increase of about 61~mK. 
The amplitude of the motional spectrum is measured for an interrogation time of 5~s. Figure \ref{fig:crystal_g2_and_spec} shows the corresponding auto-correlation of the crystal and its spectrum. Due to the large number of ions in the crystal the SNR of the measurement is still very good despite the short interrogation time and the large laser detuning.

\subsection{\textbf{$^{44}$Ca$^{+}\!+\!\,^{40}$Ca$\,\rightarrow\!\!\,^{40}$Ca$^{+}\!+\!\,^{44}$Ca charge exchange}}
\label{sec:charge_exchange}

The high resolution and short interrogation time of the secular frequency measurement technique described above makes it useful for measuring small changes in the COM-mode frequency of a large three-dimensional ion crystal. By continuously collecting auto-correlation spectra of ion crystals, we have access to a tool which can precisely detect any change in the average charge-to-mass ratio on a single particle level. In effect, this technique may be very well suited for studying cold ion-atom collisions \cite{Cote,Idziaszek,Grier} and ultra cold chemical reactions \cite{Willitsch,Gingell}.
   
In contrast to small ion stings, the COM-mode frequency change due to the crystal configuration is negligible. For an ion crystal which consists of multiple ion species the COM-mode frequency can be approximated by:
\begin{equation}
\omega_{\mbox{\tiny{COM}}}=\omega_0\sqrt{\frac{m_0}{q_0}\frac{\left\langle q\right\rangle}{\left\langle m \right\rangle}} \;,
\label{equ:secfreq}
\end{equation}
in which $q_0/m_0$ is the charge-to-mass ratio of the reference ion (0), $\left\langle q\right\rangle $ and $\left\langle m\right\rangle $ is the average charge and mass of the crystal. $\omega_0$ is the axial secular frequency of the reference ion in the trapping potential. In low energy charge exchange collisions between calcium isotopes, the electric charge of the ions will stay constant but the mass will be changed. Therefore, the COM-mode frequency can be expressed by:
\begin{equation}
\omega_{\mbox{\tiny{COM}}}=\omega_0\sqrt{\frac{N_i+N_0}{N_i \, m_i/m_0+N_0}} \;,
\label{equ:secfreq2}
\end{equation}
in which $N_i$ and $N_0$ is the ion number of isotope ($i$) and isotope ($0$) respectively. Hence, any change in the ion numbers will result in a shift of the crystal's COM-mode frequency. 

\begin{figure}[h]
\begin{center}
\includegraphics[width = 0.8 \linewidth]{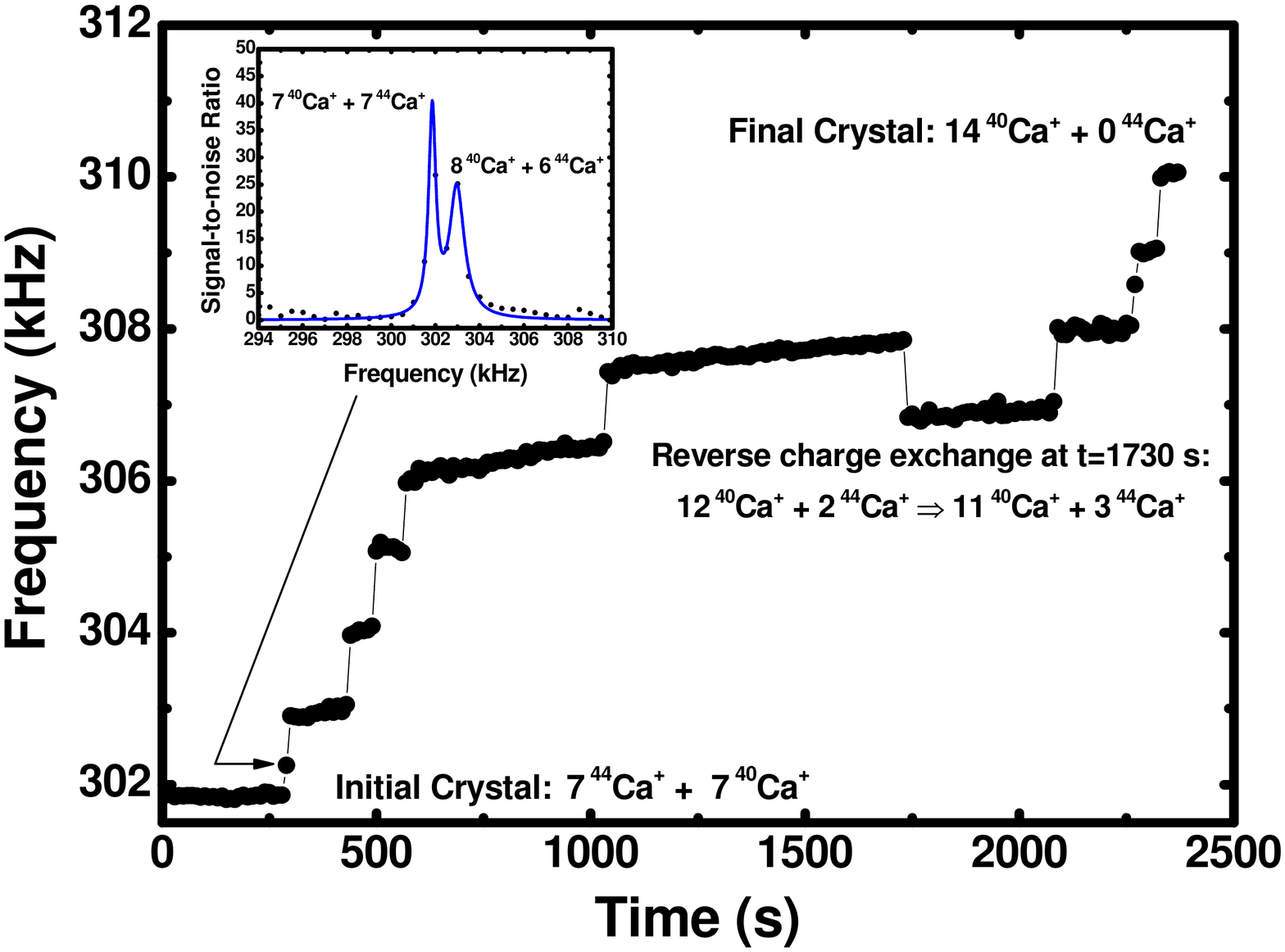}
\caption{The axial frequency of a mixed ion crystal during exposure to a neutral calcium atomic flux effusing from the oven. The motional excitation has an amplitude of 2~V, a pulse width of 2~$\mu$s and a repetition rate of 500~Hz. Auto-correlation spectra are collected continuously and each measurement has an interrogation time of 10~s. The inset shows an auto-correlation spectrum with a double peak showing the axial frequency of the crystal before and after a charge exchange event.
\label{fig:CE_experiment}}
\end{center}
\end{figure}

We employ this technique to detect the charge exchange between different calcium isotopes by measuring the axial frequency of a small mixed ion crystal exposed to a constant flux of neutral calcium atoms. The initial crystal, containing seven $^{44}$Ca-ions and seven $^{40}$Ca-ions, is created by tuning the photoionisation laser to the S$_{0}\!\!\rightarrow$P$_{1}$ transition of neutral $^{44}$Ca. Due to the overlap of the $^{40}$Ca and $^{44}$Ca resonances and the high natural abundance of $^{40}$Ca, some $^{40}$Ca-ions are inevitably created during the loading process. The initial crystal composition is determined by observing the $^{40}$Ca-ions, which are used to probe the motion of the crystal, with a CCD camera and by measuring the COM-mode frequency. From equation \ref{equ:secfreq2} the initial number of $^{44}$Ca-ions is then calculated. Switching on the oven creates a constant flux of neutral calcium at the position of the ion crystal which consists predominantly of $^{40}$Ca-atoms. Auto-correlation spectra are collected continuously with an interrogation time of 10~s. Figure \ref{fig:CE_experiment} shows the COM-mode frequency during the experiment. The steps in the frequency correspond to charge exchanges between $^{44}$Ca-ions and neutral $^{40}$Ca-atoms.  
Even though there is a change in mass of only about 4\% for each charge exchange, the change in frequency is clearly visible. Each COM-mode frequency shift occurring during the measurement is calculated and the results are listed in table \ref{tab:CE_shifts} where they are compared with the theoretical predictions. 
\begin{table}[h]
\begin{center}
\begin{tabular}{| c | c | c | c | c |}
\hline
\multicolumn{2}{|c|}{Crystal Population} & Charge & Theoretical & Measured  \\ 
\cline{1-2}
$^{40}$Ca$^{+}$ & $^{44}$Ca$^{+}$ & Exchange Time (s) & Shift (kHz) & Shift (kHz) \\
\hline \hline
8  & 6 & 300 & 1.03 & 1.08(6) \\ \hline
9  & 5 & 440 & 1.04 & 1.04(4) \\ \hline
10 & 4 & 500 & 1.06 & 1.04(5) \\ \hline
11 & 3 & 570 & 1.07 & 1.04(6) \\ \hline
12 & 2 & 1040 & 1.08 & 1.05(5) \\ \hline
11 & 3 & 1740 & 1.08 & 1.06(6) \\ \hline
12 & 2 & 2090 & 1.08 & 1.05(5) \\ \hline
13 & 1 & 2280 & 1.09 & 1.09(4) \\ \hline
14 & 0 & 2330 & 1.10 & 1.08(6) \\ \hline
\end{tabular}
\caption[]{The experimentally measured frequency shift for each charge exchange event in the measurement shown in figure \ref{fig:CE_experiment}. Theoretical shift values are calculated from equation \ref{equ:secfreq2} for the known crystal population at t=0. The exchange time is rounded to the nearest 10~s time bin.
\label{tab:CE_shifts}}
\end{center}
\end{table}

Due to the single event resolution of this technique, the time resolution can be improved beyond the interrogation time of each measurement by analysing the spectra and deducing the exact moment an event has occurred from the ratio of the amplitude of the frequency peak before and after the event (see inset in figure \ref{fig:CE_experiment}.) In addition, this method is particularly suited to detect rare events like the reverse charge exchange $^{40}$Ca$^{+}\!+\!\,^{44}$Ca$\,\rightarrow\!\!\!\,^{44}$Ca$^{+}\!+\!\,^{40}$Ca which leads to a decrease of the COM-mode frequency. This event is visible in figure \ref{fig:CE_experiment} as a decrease in COM-mode frequency at 1730~s. The plateau regions in figure \ref{fig:CE_experiment}, during which no charge exchange event occurs, are slightly inclined due to the change in temperature of the ion trap in response to the resistively heated oven which is in thermal contact with the trap mount. The observed drift in the frequency, however, is easily distinguishable from the discrete jumps of the COM-mode frequency resulting from charge exchange events.

\section{Conclusion}
\label{sec:conclusion}

We have developed a novel tool to quickly and precisely measure the secular frequency of trapped ions in rf-traps. With our set-up, we have achieved an accuracy of better than 250~Hz within an interrogation time of 100~ms and a few 10~Hz for an interrogation time of less than 30~s. This is only limited by the detection efficiency which is about 0.1\% in our set-up. By improving this, the interrogation time can be significantly reduced and/or the accuracy can be significantly increased. The heat introduced into the ions' motion is small, so that the crystal state of the ions is maintained and even configurations of small mixed ion crystals are stable during the measurement. This simple non-invasive technique is especially suited for novel applications, most notably cold chemistry with ions, high resolution spectroscopy and the measurement of the charge-to-mass ratio of ions.

We have demonstrated that this method produces accurate measurements of the COM-mode frequency of short ion strings as well as large three-dimensional ion crystals. Employing a mixed two-ion string, we were able to measure the mass of a dark ion with respect to the mass of a $^{40}$Ca-ion trapped alongside the dark ion to better than 0.08~amu. In a five-ion crystal consisting of four $^{40}$Ca-ions and one dark ion, we were able to resolve the COM-mode frequencies for the three unique configurations. The heating of the ion string due to the interrogation was low enough to maintain the configuration while still providing accurate frequency measurements. These frequencies were used to determine the mass of the dark ion. 

In order to demonstrate the potential of this method for applications in cold chemical reaction studies and charge exchange experiments, we have observed the charge exchange between two calcium isotopes. Due to the high measurement resolution, single events can be measured and from the change in the COM-mode frequency the type of process can be identified. Depending on the reaction rate and the type of reaction being studied, this method can be used to measure changes in the crystal composition for an ion number of up to a few hundred ions. Improving the fluorescence detection efficiency and increasing the axial confinement can increase this while still allowing for single event resolution. The non-invasive nature of this method makes it favorable for application in cold chemical reaction studies where a low ion temperature must be maintained \cite{Willitsch,Gingell}. For reactions in which no change in the ion crystal structure occurs, this method is ideal as it can detect such otherwise invisible changes on a single event basis. In high resolution spectroscopy of molecular ions, i.e.\ \cite{Offenberg}, the crystal composition is changed through photo-dissociation in the spectroscopy process which can be detected using the method described in this paper.

In conclusion, we have developed a novel method to measure the secular frequency of trapped ions non-invasively and with high precision which is perfectly suited to many novel applications of trapped ions.

\section*{Acknowledgements}
We thank the Ion Trap Cavity Quantum Electrodynamics Group (ITCQ) at the University of Sussex for stimulating discussions and experimental support. This work is supported by the Engineering and Physical Sciences Research Council (EPSRC) of the UK.

\end{document}